\documentclass[prb,twocolumn]{revtex4}

\usepackage{graphicx}
\begin{document}

\title{Experimental determination of the number of flux lines trapped by micro-holes in superconducting samples}

\author{A.~V.~Silhanek}
\affiliation{Nanoscale Superconductivity and Magnetism Group, Laboratory for Solid State Physics and Magnetism
K. U. Leuven\\ Celestijnenlaan 200 D, B-3001 Leuven, Belgium}

\author{S.~Raedts}
\affiliation{Nanoscale Superconductivity and Magnetism Group, Laboratory for Solid State Physics and Magnetism
K. U. Leuven\\ Celestijnenlaan 200 D, B-3001 Leuven, Belgium}

\author{M.~J.~Van Bael}
\affiliation{Nanoscale Superconductivity and Magnetism Group, Laboratory for Solid State Physics and Magnetism
K. U. Leuven\\ Celestijnenlaan 200 D, B-3001 Leuven, Belgium}

\author{V.~V.~Moshchalkov}
\affiliation{Nanoscale Superconductivity and Magnetism Group, Laboratory for Solid State Physics and Magnetism
K. U. Leuven\\ Celestijnenlaan 200 D, B-3001 Leuven, Belgium}

\date{\today}

\begin{abstract}

The influence of a periodic landscape of pinning sites on the vortex dynamics in Pb thin films is explored by ac-susceptibility measurements. For different amplitudes $h$ of the ac-drive, the ac-susceptibility $\chi=\chi^\prime+i\chi^{\prime \prime}$ exhibits a complex field dependence associated with different dynamic regimes. At very low ac-drives where both, multiquanta vortices trapped by the antidots and interstitial vortices oscillate inside the pinning potential (intra-valley motion), a small kink in $\chi^\prime(H)$ together with a very low dissipation is observed. At intermediate ac-excitations such that vortices in the antidots remain pinned whereas interstices are driven out of the pinning well, a more pronounced kink in the screening coinciding with the onset of dissipation ($\chi^{\prime \prime}(H) \neq 0$) indicates the entrance of interstitial vortices. Eventually, at high enough amplitudes all vortices are set in motion and the penetration of interstitial vortices appears as a sudden reduction of the shielding power. We show that these distinctive features allow us to clearly identify the maximum number of flux quanta per hole regardless the vortex dynamic regime.
 
\end{abstract}


\maketitle

\section{Introduction}
An increasing effort has been devoted in the last years to study the interactions of the flux line lattice with a periodic pinning landscape.\cite{martinoli78prb,silhanek-vanlook,silhanek-raedt} This problem represents a particular case of a periodic elastic medium interacting with an array of obstacles which appears in many other physical systems.\cite{runge93euro,vanblaaderen97nature,hu97prb}

It has been shown\cite{review} that the introduction of a regular array of pinning centers leads to strong commensurability effects of the vortex lattice at matching fields $H = n H_1$, where $H_1 \approx \Phi_0/d^2$, $\Phi_0$ is the superconducting flux quantum and $d$ is the period of the pinning array. These effects manifest themselves in different ways depending on the degree of stability of the vortex configuration and the nature of the vortex pinning. For fields $H < H_s = n_s H_1$, where $n_s$ is the maximum number of flux quanta that a pinning center can trap (the saturation number), every multiquanta $n\Phi_0$-vortex ($n<n_s$) is strongly pinned by an individual defect. A particular case arises at $H = n_s H_1$, commonly described as a Mott insulator phase with a large compressional elastic modulus $C_{11}$ which may give rise to a fixed density of vortices over a finite range of $H$.\cite{nel-vin,zhukov-condmat} 

A different situation occurs for $H > n_s H_1$ where fully occupied pinning centers are no longer attractive potentials but interact repulsively with the incoming vortices. In contrast to the localized Mott or insulating-like phase, in this field range, a metallic-like behavior is observed due to the weakly pinned interstitial $\Phi_0$-vortices. The peculiar transition at $H = n_s H_1$, from localized to delocalized particles is characterized by a sudden reduction of the depinning force density together with a substantial increase in the average vortex mobility. For $H > n_s H_1$, commensurability effects between the interstitial vortex configuration and the pinning array, give rise to local increments of the critical current $J_c$.\cite{martin97,metlushko-99b,vanlook02}

%ac-susceptibility
Although these effects have been mainly studied by transport and dc-magnetization measurements, less explored has been the ac response of these systems.\cite{metlushko-98,metlushko-99a,metlushko-99b,delong-02} Ac-susceptibility allows one to readily cover a wide range of dynamic regimes from a passive linear regime, preserving the initial vortex configuration, to the highly disturbing critical state regime, by simply sweeping the amplitude $h$ of the ac-drive.\cite{review-nos}

%in this work
In the present work we study the field and temperature dependence of the vortex pinning in Pb thin films with a square array of micro-holes by means of ac-susceptibility $\chi=\chi^\prime+i\chi^{\prime \prime}$. We found that $\chi(H)$ exhibits a complex behavior associated with a rich variety of vortex dynamics phases. The possibility to access different regimes by tuning the amplitude of the drive field enables us to determine the saturation number $n_s$ for several temperatures $T$ and dc-fields $H$, which is difficult to achieve by other techniques.

\section{Experimental Details}
%EXPERIMENTAL
The experiments were conducted on two Pb thin films with a square antidot array of period $d=1.5~\mu$m, which corresponds to a first matching field $H_1=9.2$ G. The antidots have a square shape with a size $b=0.8~\mu$m. The tag, thickness $\delta$ and critical temperature of the used samples are, AD100 ($\delta=100$ nm, $T_c=7.22$ K) and AD65 ($\delta=65$ nm, $T_c=7.21$ K), respectively. From the $T_c(H)$ slope we have estimated, for both samples, a superconducting coherence length $\xi(0) \sim 33 \pm 1$ nm. 

It is important to mention that even though bulk Pb is a Type-I superconductor with a Ginzburg-Landau parameter\cite{poole} $\kappa = 0.45 < 1/\sqrt{2}$, thin films of this material become Type-II superconductors below a certain crititcal thickness $\delta_c$.\cite{tinkham} In the particular case of Pb films the critical thickness\cite{dolan,rodewald} is $\delta_c \sim$ 250 nm and therefore, both samples used in the present work exhibit a clear mixed state characteristic of the Type-II superconductors. In addition, it has been recently shown that this effect becomes more pronounced in patterned samples where the effective penetration depth is further increased due to the presence of the nano-structuring.\cite{wahl}

The predefined resist-dot patterns were prepared by electron-beam lithography in a polymethyl metacrylate/methyl metacrylate (PMMA/MMA) resist bilayer covering the SiO$_2$ substrate. A Ge(20~\AA)/Pb/Ge(200~\AA) film was then electron-beam evaporated onto this mask while keeping the substrate at liquid nitrogen temperature. Finally, the resist was removed in a lift-off procedure in warm acetone. 

The ac-measurements were carried out in a commercial Quantum Design-PPMS device with drive field amplitudes $h$ ranging from 2 mOe to 10 Oe, and the frequency $f$ from 10 Hz to 15 kHz. In this frequency window we have found that $\chi$ depends only weakly on $f$ and therefore we report results obtained at the same frequency $f=3837$ Hz. In all cases, the data were normalized to have a total step $\Delta \chi^{\prime} =$ 1, with $H=$0 at $T < 5$ K.

%RESULTS AND DISCUSSION
\section{Results and Discussion}
\begin{figure}[htb]
\centering
\includegraphics[angle=0,width=80mm]{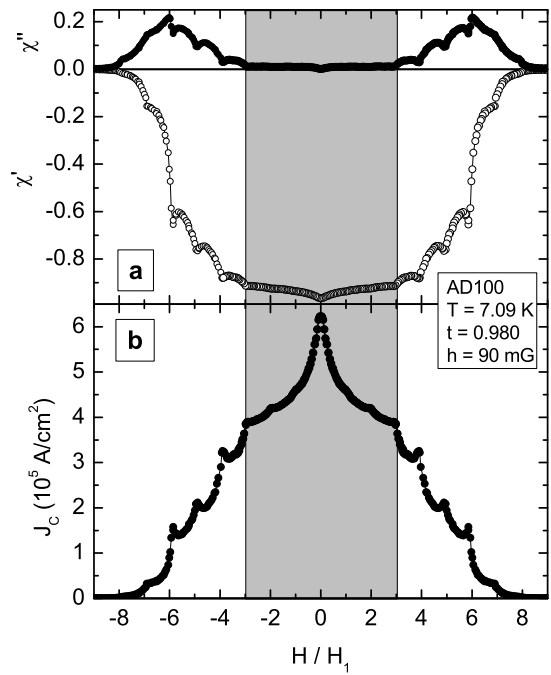}
\caption[]{{\small (a) Field dependence of the real $\chi^\prime$ (open symbols) and imaginary $\chi^{\prime \prime}$ (solid symbols) components of the ac-susceptibility for the AD100 sample at $h=90$ mG. (b) Field dependence of the critical current extracted from the $\chi^\prime(H)$ curve shown in (a) (see text). The shadowed area indicates the multiquanta vortex regime.}}
\label{fig1}
\end{figure}

As it has been demonstrated in previous theoretical\cite{reichhardt} and experimental studies,\cite{martin97,metlushko-99b,vanlook02} rearrangements of the vortex lattice at the matching fields $n H_1$ have a profound effect on the critical current. Accordingly, these changes in the pinning properties of the flux line lattice should be also reflected in changes of the efficiency to screen out a perturbative external ac-field.\cite{metlushko-99b} In order to corroborate this effect we have measured $\chi$ as a function of $H$ at fixed $T$ and $h$. Figure \ref{fig1}(a) shows the $\chi^\prime(H)$ (open circles) and $\chi^{\prime \prime}(H)$ (filled circles) dependencies for the AD100 sample at $T =$ 7.09 K and $h =$ 90 mG.\cite{note}

At low $H$, a smooth decrease of the screening together with a very low dissipation $\chi^{\prime \prime} \sim $ 0, is observed as $H$ is increased. At $H=3H_1$, a well defined kink in $\chi^\prime(H)$  coinciding with the onset of dissipation, occurs. This behavior indicates that at $H = 3 H_1$ a substantial change in the pinning nature takes place. We argue that at this field it is no longer energetically convenient for incoming vortices to sit into the antidots and consequently, interstitial vortices appear. According to this scenario, although for $H < 3 H_1$ the current induced by the ac-drive is not large enough to move out of the potential well vortices pinned by the antidots, for $H > 3 H_1$ the same force can easily set in motion the weaker pinned interstitial vortices giving rise to a clear change in the screening properties. This behavior is consistent with recent transport measurements conducted on similar samples.\cite{silhanek-vanlook} In brief, the sudden reduction of the effective pinning force density at $H=3H_1$ allows for the used ac-amplitude to connect the ascending and descending branches of the dc-hysteresis loop giving rise to a detectable dissipation.\cite{morozov} For lower dc-fields, the minor loops built with the ac-cycle do not bridge the magnetization branches and the dissipation is negligible.

For fields $3H_1 < H < 6H_1$ a series of peaks in the screening are seen at each matching field, in agreement with the experimentally observed local enhancement of the critical current.\cite{martin97,metlushko-99b,vanlook02} These maxima in the screening are accompanied by local dips in the dissipation. Interestingly, at $6H_1$ the screening undergoes a dramatic reduction, which coincides with the maximum dissipation. For higher fields, small peaks are still visible at $7H_1$ and $8H_1$, before the system enters the normal state. Now, these small peaks correspond to local maxima in the dissipation.

As we will demonstrate below, the observed sudden reduction at $6H_1$ can be an artifact owing to the higher sensitivity of the screening in that particular range of ac-penetration depths ($\Lambda$) and it is not necessarily related with a change in the vortex dynamic regime. Indeed, for the sake of clarity consider for instance that we can approximate our sample shape by a strip with both, $H$ and $h$, applied perpendicular to the plane of the strip. In this case, according to the Bean critical state model, the components of the ac-susceptibility are given by,\cite{herzog}

\begin{eqnarray}
&\chi^\prime=-\frac{Tanh(s)}{s},\label{eq1}\\
&\chi^{\prime\prime}=-\frac{Tanh(s)}{s}+\frac{Tanh(s/2)}{s/2},\label{eq2}
\end{eqnarray}
where $s \approx \Lambda/\delta  \propto J_c^{-1}$. The derivative of the eq.(\ref{eq1}) and therefore the sensitivity of $\chi^\prime$ to small changes in $\Lambda$, maximizes at $s=1$, i.e. when the Bean penetration depth coincides with the sample thickness. On the other hand, the situation $\Lambda \sim \delta$ is achieved at the maximum of the dissipation $\chi^{\prime \prime}$,\cite{clem-san} thus naturally accounting for the systematic coincidence of the field position of the maximum dissipation and the jump in the screening. Furthermore, using the strip geometry approximation we can estimate from $\chi(H)$ the field dependence of the critical current $J_c$. The result of this analysis is shown in Figure \ref{fig1}(b). In this Figure we observe that the height of the jump at $6H_1$ has now no particularities and is similar to that observed at $4H_1$. In contrast to that, the distinct feature signaling the entrance of interstitial vortices at $3H_1$ is still clearly seen.

\begin{figure}[htb]
\centering
\includegraphics[angle=0,width=80mm]{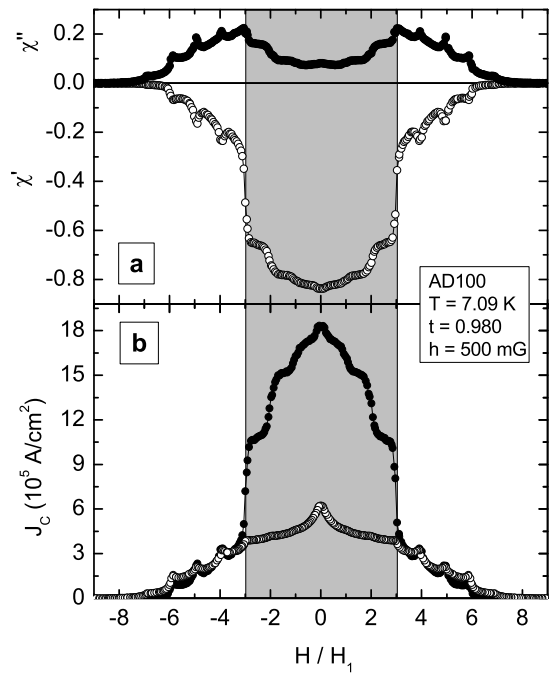}
\caption[]{{\small (a) Field dependence of the real $\chi^\prime$ (open symbols) and imaginary $\chi^{\prime \prime}$ (solid symbols) components of the ac-susceptibility for the AD100 sample for $h=500$ mG, at $t=0.980$. (b) Critical current as a function of field, obtained from the $\chi^\prime(H)$ curves for $h=90$ mG and $h=500$ mG, at $t=0.980$ (see text). The shadowed area indicates the multiquanta vortex regime.}}
\label{fig2}
\end{figure}

It is important to note that the kink in $\chi^\prime(H)$ at $H=3H_1$ allows us to determine the saturation number $n_s(T,b,\xi)$, which is an intrinsic property of the pinning array and therefore it is independent of $h$. In contrast to that, the big jump associated with the peak in the dissipation, should occur at a field $H$ such that $J_c(H) \approx h/\delta$ and thus should depend on the strength of the ac-drive $h$. This is in fact confirmed by the $\chi(H)$ measurements performed on the same sample at the same temperature, with a higher excitation $h = 500$ mG (see Figure \ref{fig2}(a)). In this case, we observe that the largest reduction of the screening is shifted down to $H_s = 3 H_1$ following the position of the maximum dissipation. 

%Additionaly, small peaks at $H_1$ and $H_2$ as a result of the long range vortex-vortex interaction, become more evident.

Performing the same procedure described above, we estimate the critical current $J_c(H)$ obtained from the $\chi^\prime(H)$ curve at $h=500$ mG (see filled circles in Figure \ref{fig2}(b)). Now we can see that after this transformation the sharp change at $H_s = 3 H_1$ persists and is not an artifact due to the amplification in the screening sensitivity. Since, according to the Bean model $J_c$ is independent of the used ac-excitation, the critical currents obtained for $h=90$ mG and $h=500$ mG should coincide in the field range where this approximation holds. This behavior becomes apparent in Fig.\ref{fig2}(b) where we superimpose the $J_c(H)$ curve for $h=90$ mG, already shown in Fig.\ref{fig1}(b) (open circles). The remarkable coincidence of the data for fields $H>H_s$ indicates that in this particular field range the simple Bean model is a very robust approximation and correctly accounts for the observed behavior. On the other hand, the difference observed at $H<H_s$ is due to the fact that for $h=90$ mG the used Bean critical state is no longer valid and therefore the $J_c(H)$ curve so determined is ill-defined.

Another feature shown in Fig.\ref{fig2} is that for $H < H_s$, where the antidots act as attractive centers, commensurability effects manifest themselves as small reductions in the screening as $H$ increases. In this field range, the vortex-antidot interaction dominates over vortex-vortex interaction and thus, the decrease in the effective pinning potential strength at $n H_1$ will result in a reduction in the shielding power. For $H > H_s$, filled antidots become repulsive centers and an opposite behavior is observed.\cite{moshchalkov96} At these high fields, vortex-vortex interactions are more relevant and consequently, collective effects give rise to local increments in the screening properties at every commensurability field. We have verified that these features signaling the transition from attractive to repulsive vortex-pin site interactions which allow one to identify the saturation number $n_s$, were present in six other samples with periodic pinning array.

Let us now analyze the experimentally obtained $n_s = 3$ value. According to the theoretical estimation of Mkrtchyan and Schmidt (MS),\cite{schmidt} the number of trapped vortices by a hole is $n_s \approx b/4\xi(T)$ which in our particular case gives $n_s \sim 1$. Although this value is smaller than the experimentally determined above, it is important to note that this model underestimates the real $n_s$ value since it considers a single cylindrical cavity with radius $b \ll \lambda$. Clearly, this hypothesis is not fulfilled in our system where $b > 2 \lambda(T)$. A more detailed analisys of the validity of this model could be obtained from the temperature dependence $n_s(T)$. However, matching effects reported here are restricted to a very narrow temperature range\cite{note} close to $T_c$ where $n_s$ remains almost constant and no evident temperature dependence is observed. Additionally, it is also expected that $n_s$ increases as the applied dc-field is increased.\cite{baert,doria,buzdin} An extension of the original work of Mkrtchyan and Schmidt to arbitrary large cavity radius has been recently done by  Nordborg and Vinokur\cite{nordborg} using the London approximation. Addiotonally, by simple energetic considerations Buzdin\cite{buzdin} showed that in a triangular vortex lattice, two-quanta vortex becomes energetically favorable for temperatures such that $b^3<\xi(T)\lambda(T)^2$, a condition that, in our sample, is satisfied for $t<0.995$. The experimental evidence on this point also suggests that a higher filling than that predicted by MS has to be considered. Indeed, Bitter decoration experiments performed by Bezryadin et al.\cite{pannetier} on a Nb film with a triangular array (period $d=6.1 \mu$m) of blind holes (size $b=0.8 \mu$m), showed that $n_s=3$ very close to the critical temperature for $H=6.37$ G. More recently, Grigorenko et al.\cite{grigorenko} using scanning Hall probe microscopy in a Pb film with a square array of antidots of period $1.5 \mu$m and hole size $b= 0.66 \mu$m showed that $n_s=2$ at $t=0.77$. Although this number is somehow smaller than our estimate, it should be noted that in those experiments, the measurements are performed in a field-cooled metastable state which corresponds to a frozen vortex configuration at a higher temperature. On top of that, since the $n_s(b)$ is a stepwise increasing function,\cite{pannetier} even though the holes used in their work are slightly smaller than ours, this small difference can easily lead to an extra vortex trapped per hole. Thus, in general, our determination of the saturation number by means of ac-susceptibility is consistent with the previously reported results using alternative techniques. 

\begin{figure}[htb]
\centering
\includegraphics[angle=0,width=80mm]{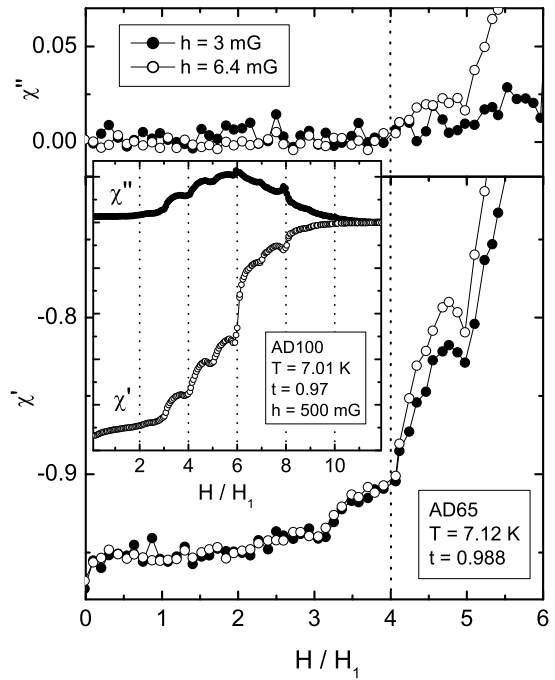}
\caption[]{{\small Main panel: $\chi(H/H_1)$ for the AD65 sample for $h=3$ mG and $h=6.4$ mG, at $t=0.988$. The dotted line indicates the onset of dissipation. Inset: $\chi(H/H_1)$ for the AD100 sample at $h=500$ mG and $t=0.97$.}}
\label{fig3}
\end{figure}

So far we have shown that at low $h$ amplitudes, the current induced by the ac-drive is not large enough to move vortices trapped by antidots out of the pinning potential but can set in motion the weakly pinned interstitial vortices. According to this picture, it should be possible to reach a situation where interstitial vortices perform intra-valley oscillations as well, by further decreasing the ac-drive. To check this possibility we have measured the ac response of the sample AD65 at very low amplitudes ($h=3$ mG and $h=6.4$ mG), as shown in the main panel of Figure~{\ref{fig3}}. The dotted line indicates the upper limit of the field range ($H < 4H_1$) where the detected dissipation lies below our experimental resolution ($\chi^{\prime \prime} \sim 0$) and therefore where vortices oscillate inside the pinning potential. Within this regime the ac-response is independent of $h$, confirming that the system is in the linear regime. We also note that a clear kink at $H = 3H_1$ signals the transition from a intra-valley motion of multiquanta vortices trapped by the antidots, to an intra-valley motion of interstitial vortices inside the potential produced by the strongly pinned neighbors and vortices trapped by the holes. For fields $H > 4H_1$ a finite dissipation together with an amplitude dependent response indicate the onset of inter-valley motion of interstitial vortices, whereas vortices in the antidots remain pinned.

We turn now to the analysis of the field dependence of dissipation  $\chi^{\prime \prime}$. As we noticed previously, for fields smaller than that where the maximum dissipation takes place, the local maxima in the screening, and therefore in the critical current, are accompanied by local dips in the dissipation. In contrast, for higher fields, peaks in $|\chi^\prime|$ give rise to local increments of $\chi^{\prime \prime}$. This effect is more clearly seen in the inset of Fig.\ref{fig3}, where we show $\chi(H)$ for the AD100 sample at $t=0.97$ and $h=500$ mG. The observed behavior can be easily understood within a critical state scenario. Indeed, in the simplest case of a Bean critical state model for a thin sample with applied field $H$ perpendicular to the plane of the film (see eq.(\ref{eq1}) and eq.(\ref{eq2})), the maximum dissipation occurs when $h$ reaches the full penetration field $H_d(T,H)=J_c \delta$.\cite{herzog,clem-san} According to this model, the asymptotic limits of the real and imaginary ac-susceptibility components are related by,

\begin{eqnarray}
\label{eq:1}
&&\chi^{\prime \prime} \approx \frac{3}{4}\left(1-|\chi^\prime| \right),~~~~~ h \ll J_c \delta  \\
&&\chi^{\prime \prime} \approx |\chi^\prime|,~~~~~ h \gg J_c \delta \nonumber
\end{eqnarray}

From these equations we can see that for low dc-fields ($h \ll J_c \delta$) an enhancement of the screening leads to a suppression of the dissipation, whereas an opposite behavior is predicted for high fields ($h \gg J_c \delta$), in agreement with our experimental observation. It is interesting to note that in this case the commensurability effects have the benefit of acting as markers in $J(H)$ which eventually help to study and identify the different dynamic regimes. 

%deppining of multiquanta
Finally we would like to note that no clear indications of the depinning of vortices trapped by the antidots have been observed. This depinning process of multiquanta vortices at $H_t$ represents a peculiar transition from a state where supercurrents flow around an antidot (for $H < H_t$) with the $n\Phi_0$-vortex to a $n$ depinned $\Phi_0$-vortices state (for $H > H_t$). This metamorphoses of single to multiple entities is a very interesting issue that has not been addressed so far and clearly deserves further theoretical and experimental investigations.

%Indeed, since a $n\Phi_0$ multiquanta vortex is not energetically stable outside of the hole, once the induced currents surpass the depinning threshold a transformation into $n$ single quantum vortices occurs. After this fragmentation, a retrapping of vortices might take place leading to a complex backward and forward re-transformation. The description of the multivortex depinning can be substantially modified if the recently proposed Priour and Fertig's approach\cite{priour} is considered. In the case of a single vortex, these authors showed that instead of the assumed rigid core depinning, the driving current elongates the vortex core which can eventually reach the neighbor pinning site thus allowing the vortex current to hop from site to site. Within this picture interstitial vortices resulting from a depinning process may never form, in contrast to what is commonly assumed. Less clear is the extension of this analysis to consider the case of multiquanta vortex depinning, where ``vortex fragmentation" can appear. 

\section{Conclusion}
%summary
In summary we have shown that the field dependence of the ac response of samples with a periodic pinning array exhibits a rich behavior associated with different dynamic regimes. In particular, we show that for low drive fields two distinctive features emerge. First, at low dc-fields $H$, a kink in the screening $\chi^\prime(H)$ indicates the entrance of interstitial vortices in the sample. This provides a reliable determination of the saturation number $n_s$ of the pinning structure which is hard to find by other indirect methods. Second, at a higher $H$ the enhanced sensitivity of the ac-screening, when the penetration depth $\Lambda \sim t$, accents the drop on the critical current leading to a dramatic reduction of $\chi^\prime(H)$. In general, this analysis represents a powerful method to study the strength of different pinning centers arranged periodically and in particular, provides an alternative way for determining the maximum number of flux lines trapped per pinning site as a function of temperature. 

\acknowledgments
%\section*{Acknowledgements}
We would like to thank R. Jonckheere for fabrication of the resist pattern and L. Van Look for helpful discussions. This work was supported by the Belgian Interuniversity Attraction Poles (IUAP), Research Fund K.U.Leuven GOA/2004/02, the Fund for Scientific Research Flanders (FWO) and ESF ``VORTEX'' program. MJVB is a postdoctoral Research Fellow of the FWO.

\bibliographystyle{prsty}

\begin{thebibliography}{10}
\bibitem{note}
Since matching features are only present at temperatures very close to $T_c$, we have restricted the range of shown data to $t=T/T_c>0.97$. In the used samples we found that for $t<0.97$ matching effects progressively fade out.
\end{thebibliography}

\end{document}